\documentstyle[graphicx,referee]{mn}
\newcommand{\ha}{\ifmmode {\rm H}\alpha \else H$\alpha$\fi}
\newcommand{\hb}{\ifmmode {\rm H}\beta \else H$\beta$\fi}
\newcommand{\lya}{\ifmmode {\rm Ly}\alpha \else Ly$\alpha$\fi}
\newcommand{\lyb}{\ifmmode {\rm Ly}\beta \else Ly$\beta$\fi}
\newcommand{\ciii}{\ifmmode {\rm C}\,{\sc iii} \else C\,{\sc iii}\fi}
\newcommand{\civ}{\ifmmode {\rm C}\,{\sc iv} \else C\,{\sc iv}\fi}

\title{Ly$\alpha$~Polarisation in Thick H I Media}
\author[Lee and Ahn]{Hee-Won Lee$^1$ and Sang-Hyeon Ahn$^2$
\\
$^1$ Department of Geoinformation Sciences, Sejong University, Seoul 143-747, 
Korea
\\
$^2$ Korea Institute for Advanced Study, 207-43 Cheongyangri-dong, 
Dongdaemun-gu, Seoul 130-012, Korea}

\date{2002 March 4}

\begin{document}

\maketitle

\begin{abstract}
We have investigated the linear polarisation of Ly$\alpha$
transferred in a static, dustless, and thick plane-parallel slab that is
uniformly filled with neutral hydrogen. 
Because the scattering phase function is characterised by the dipole
type for wing scatterings of Ly$\alpha$, which is in high contrast 
with more isotropic scattering phase function associated with resonant 
scatterings, the polarisation of the emergent Ly$\alpha$ is dominantly 
controlled by the number of wing scatterings before escape from the slab. 
When the scattering medium is extremely thick, photons emergent
in the grazing direction are polarised in the direction perpendicular 
to the slab normal up to $12\%$, which coincides with the limit attained by
the Thomson scattered radiation in a semi-infinite thick medium investigated 
by Chandrasekhar in 1960. The number of wing scatterings before escape
is sensitively dependent on the product of the Voigt parameter $a$
and the line-centre optical depth $\tau_0$, with which the variation of 
the degree of polarisation can also be explained in comparison with 
the Thomson scattering case. The polarisation direction is parallel to 
the slab normal when $a\tau_0<10^3$, while it is perpendicular when 
$a\tau_0>10^3$.
Our simulated spectropolarimetry of Ly$\alpha$ from a highly thick medium
is characterised by negligible polarisation around the line centre, 
parallel polarisation in the near wing parts and perpendicular polarisation 
in the far wing parts, where the polarisation direction is measured
with respect to the slab normal.
The polarisation flip naturally reflects the diffusive nature of the
Ly$\alpha$ transfer process in both real space and frequency space.
We also discuss the beaming effect that is associated with
the parallel polarisation in the near wing parts of the emission.
\end{abstract}


\section{Introduction}

In the ultraviolet band there are many resonance lines
arising from the ${\rm S}_{1/2}\rightarrow {\rm P}_{1/2, 3/2}$
transitions including CIV$\lambda\lambda 1548,1551$,
NV$\lambda\lambda 1238,1241$, MgII$\lambda\lambda 2798,2800$,
and Ly$\alpha \lambda1216$.
These lines are found to be prominent and conspicuous in many line-emitting 
objects including quasars, starburst galaxies, and planetary nebulae 
because they are major coolants in plasmas with temperature 
$T\sim 10^4-10^5{\rm\ K}$. The atomic physics associated with 
the ${\rm S}_{1/2}\rightarrow {\rm P}_{1/2, 3/2}$ transitions
provides a nice and simple example of the quantum interferences between 
the excited levels.

It is well-known that the resonance scattering
between two levels with $J=1/2$ yields isotropic outgoing 
radiation that is completely unpolarised.
This is in high contrast with the fact that a $90^\circ$ resonance
scattering of an unpolarised photon associated with a 
$J=1/2\rightarrow J=3/2$ transition
yields the maximum degree of polarisation $3/7$ (e.g. Stenflo 1980;
Lee, Blandford \& Western 1994). 
Stenflo (1976a, 1976b, 1980, 1994, 1996) and Lee \& Blandford (1994)
studied the polarisation of resonance lines. 

Interesting phenomena occur when the
incident radiation is scattered in off-resonance regimes,
where both the polarisation and the angular distribution of 
the outgoing radiation are sensitively dependent on the wavelength 
of the incident radiation, as is illustrated by Stenflo (1980).
One general result is that scattering in far off-resonance is 
characterised by the classical Rayleigh scattering phase function.

Being the most prominent emission features with a very small 
fine-structure level-splitting in the excited states of hydrogen,
Ly$\alpha$ should be treated with more emphasis and care.
Due to the small fine-structure level-splitting of hydrogen,
wing scatterings of Ly$\alpha$ are characterised
by the classical Rayleigh scattering phase function,
which is the same phase function associated with the Thomson scattering. 
Hence, when wing scatterings of Ly$\alpha$
become important in a thick medium of hydrogen, we naturally expect that
the Ly$\alpha$ line transfer may be described in a  way similar to
the transfer of continuum photons in a Thomson scattering medium.

Chandrasekhar (1960) investigated the polarisation
of photons transferred in a very thick electron gas,
where he showed the degree of polarisation reaches 11.7\%
for photons emergent in the grazing direction of very thick 
and plane-parallel media.
Phillips \& M\'esz\'aros (1986) also investigated in a numerical way
the radiative transfer in electron clouds, for both optically thin
and thick cases.
They found that the photons are beamed to the direction normal
to the slab, and obtained 11.7\% of polarisation confirming the work
of Chandrasekhar (1960).

The vertical column density $N_{\rm HI}$ of normal galaxies
and dwarf galaxies ranges $N_{\rm HI}=10^{19}-10^{22} {\rm cm}^{-2}$.
The line-centre optical depth is related with the HI column 
density $N_{\rm HI}$ by 
\begin{equation}
\tau_0 \equiv 1.41\ T_{4}^{-1/2}
\left[{N_{\rm HI} \over {10^{13} \rm cm^{-2}}}\right],
\end{equation}
where $T_4$ is the temperature of the medium in units of $10^4 {\rm K}$.
This implies that the Ly$\alpha$ line-centre optical depth
$\tau_0=10^6-10^9$ for most galaxies, and therefore one
needs to deal with an extremely thick medium. It appears that many
galaxies found in the early universe exhibit asymmetric Ly$\alpha$ emission,
which is supposed to be emergent from media with these high optical depths.
Therefore, if the scattering medium possesses any geometrical or
kinematical anisotropy, we may expect that the Ly$\alpha$ emission 
from these objects will be polarised. 

A preliminary investigation of the polarisation of Ly$\alpha$ emission 
in a thick medium has been presented in Lee \& Ahn (1998).
They considered an anisotropically expanding and 
optically thick neutral medium with a Hubble-type flow,
and found that the emergent Ly$\alpha$ photons 
are linearly polarised with a degree of polarisation up to 10\%
in the direction parallel to the slab normal.
They also applied their calculation to a hemispherical shell 
partly obscured by an opaque component such as a disc, 
and showed that about 5\% of polarisation may develop.

In this paper, we calculate the polarisation of Ly$\alpha$
in more detail with the particular attention on the polarisation direction
as a function of the optical depth.  Section 2
describes the configuration of our model and the Monte Carlo code.
In section 3, we show the results, and
explain microscopic processes of scatterings that lead to 
the development of Ly$\alpha$ polarisation, which is followed by
the summary section.

\section{Model}

In a very thick medium, wing scattering plays a very important role
in the transfer of resonance line photons, which is
similar to a diffusion process. 
Previous researchers (Avery \& House 1968; Adams 1972;
Harrington 1973; Neufeld 1990) introduced the diffusion approximation
where only wing scatterings take place during the transfer.
This approximation is very efficient for the extremely thick cases
with $a\tau_0>10^3$, where $a$ is the Voigt parameter defined as
$a\equiv \Gamma/4\pi\Delta\nu_D$, the ratio of the natural width $\Gamma$
and the Doppler frequency width $\Delta\nu_D=\nu_0(v_{th}/c)$ with
$v_{th}$ and $\nu_0$ being the thermal speed and the line centre
frequency, respectively. We introduce the dimensionless
parameter $x$ that measures the frequency shift from the line centre
in units of $\Delta\nu_D$,
\begin{equation}
x\equiv (\nu-\nu_0)/\Delta\nu_D.
\end{equation}

However, Ly$\alpha$ photons experience both core scatterings and
wing scatterings and may alternate between these two types of scattering
during the line transfer. A large number of
core scatterings change the polarisation vector in a random way 
yielding an isotropic and unpolarised radiation field.
Therefore, we have to consider not only wing scatterings but also
core scatterings in order to compute accurately the polarisation
of Ly$\alpha$ emergent from a very thick H~I medium.

Ahn, Lee, \& Lee (2000, 2001, 2002) developed a Monte Carlo code
which can deal with the Ly$\alpha$ line transfer in thick media.
They first dealt with moderate optical depths, because the diffusion 
approximation is not valid in this optical depth regime.
They developed an accelerating scheme for the code
in order to improve the computing speed, which enabled one to
calculate the line transfer in extremely thick media.
The code also accurately treats the partial frequency redistribution
and is equipped with a subroutine that differentiates
between resonant core scatterings and non-resonant wing scatterings,
providing different scattering phase functions according to the type
of scatterings.  

In this work, we use the same Monte Carlo code, of which the description
can be found in our previous work. 
Our code treats the scattering of Ly$\alpha$ photons
in a manner very faithful to the atomic physics associated
with the fine-structure of hydrogen.  The level-splitting
of $2{\rm P}_{{1\over 2},{3\over 2}}$, the excited state of Ly$\alpha$
transition, is $10{\rm GHz}$, which amounts
to the Doppler width of $1.34{\rm\ km\ s^{-1}}$ relative to Ly$\alpha$.
Because this is much smaller than the thermal speed 
of a medium with $T\ge 100{\rm\ K}$,
one normally neglects the $2{\rm P}_{{1\over 2},{3\over 2}}$ level-splitting,
treating it as a single level.
Even when $T \le 100{\rm\ K}$, the level-splitting effects may be neglected
in the cases where the line centre optical depth of the scattering
medium is very large. This is because the escape of line photons
can only be made in considerably far wings where the level-splitting
is negligible.

Since we are particularly interested in the development of polarisation 
of Ly$\alpha$, we provide the detailed description of the code 
about the angular redistribution and the polarisation of the scattered 
radiation. We adopt the density matrix formalism to describe the angular 
distribution and polarisation of the scattered Ly$\alpha$, 
where the density operator is represented by a $2\times2$ 
Hermitian matrix $\rho$. The usual relation with
the conventional Stokes parameters $I, Q, U$ and $V$ is given by
\begin{eqnarray}
\rho_{11} &=& {I+Q\over 2}, \quad \rho_{22} = {I-Q\over 2} \nonumber \\
\rho_{12} &=& {U+iV\over 2}, \quad  \rho_{21} = \rho_{12}^* 
\end{eqnarray}

Fig.~1 shows the model configuration adopted in this paper.
We consider a static, uniform, and optically thick HI slab
with the total optical thickness $2\tau_0$ along the $z-$axis
that is chosen to be the slab normal. 
The completely unpolarised Ly$\alpha$ source is located at the origin of 
the coordinate system and $\mu$ is defined as the cosine of the angle 
between the wavevector of the emergent photon and the slab normal.
We will not consider the circular polarisation because of the azymuthal 
symmetry assumed in this work. Therefore $V=0$ and $\rho$ may be regarded
as a real matrix. 

In our Monte Carlo code, we only consider three types of scattering of 
Ly$\alpha$, which are 
resonance scattering with ${\rm S}_{1/2}\rightarrow {\rm P}_{1/2}$,
resonance scattering with ${\rm S}_{1/2}\rightarrow {\rm P}_{3/2}$ 
and wing scattering far from both resonance transitions.
At each scattering event, we compute the probability which type
of scattering has occurred. The procedure for the computation of this
probability is described in detail in our previous work (Ahn et al. 2000).

The incident radiation is described by the density matrix $\rho$ 
and the wavevector $\hat{\bf k}
=(\sin\theta\cos\phi,\sin\theta\sin\phi, \cos\theta)$, and the scattered
radiation is described by the corresponding primed quantities $\rho'$
and 
\begin{eqnarray}
\hat{\bf k'}=(\sin\theta'\cos\phi',\sin\theta'\sin\phi', \cos\theta').
\nonumber
\end{eqnarray}
We choose the polarisation vectors 
\begin{eqnarray}
{\bf\epsilon_1}  &=& (-\sin\phi,\cos\phi,0) \nonumber \\
{\bf \epsilon_2} &=& (\cos\theta\cos\phi,\cos\theta\sin\phi, -\sin\theta)
\end{eqnarray}
for an incident photon and correspondingly ${\bf\epsilon_1'}, {\bf\epsilon_2'}$
for the scattered photon. 

With this choice of the polarisation basis
and the scattering geometry shown in Fig.~1, the polarisation will develop
either in the direction parallel to the slab normal (the ${\bf\epsilon_2}$
component) or in the direction perpendicular to the slab normal
(the ${\bf\epsilon_1}$component).  Therefore, we may expect that 
the polarisation of the emergent radiation will be described completely 
by $P=\rho_{11}-\rho_{22}$, the difference of the diagonal elements of 
the density matrix, which we will call the degree of polarisation. Hence
in this convention, the positive degree of polarisation means polarisation
in the direction perpendicular to the slab normal, and the negative 
polarisation is polarisation in the direction parallel to the slab normal.
In this work, we refer the polarisation direction with respect to the
slab normal vector which is the symmetry axis of the scattering medium.

When a wing scattering occurs, the density matrix associated with
the scattered radiation is given by
\begin{eqnarray}
\rho_{11}' &=&\rho_{11}\cos^2\Delta\phi)
-\rho_{12}\cos\theta\sin2\Delta\phi) \nonumber \\
&&+\rho_{22}\sin^2\Delta\phi\cos^2\theta  \nonumber  \\
\rho_{12}' &=& {1\over2}\rho_{11}\cos\theta'\sin2\Delta\phi
+ \rho_{12}(\cos\theta\cos\theta'\cos2\Delta\phi \nonumber \\
&&+\sin\theta\sin\theta' \cos\Delta\phi \\
&&+\rho_{22}\cos\theta(-\sin\theta\sin\theta'\sin\Delta\phi \nonumber\\
&&-{1\over2}\cos\theta\cos\theta'\sin2\Delta\phi) \nonumber\\
\rho_{22}' &=& \rho_{11}\cos^2\theta'\sin^2\Delta\phi \nonumber \\
&&+\rho_{12}\cos\theta'(2\sin\theta\sin\theta'\sin\Delta\phi \nonumber\\
&&+ \cos\theta\cos\theta'\sin2\Delta\phi) \nonumber \\
&&+\rho_{22}(\cos\theta\cos\theta'\cos\Delta\phi+\sin\theta\sin\theta')^2,
\nonumber
\end{eqnarray}
where $\Delta\phi=\phi'-\phi$.

When a given scattering is a resonance scattering associated with the
transition ${\rm S}_{1/2}\rightarrow {\rm P}_{3/2}$, 
then the density matrix is obtained by the following relation
\begin{eqnarray}
\rho_{11}' &=& \rho_{11}(5+3\cos2\Delta\phi) \nonumber \\
&&+ \rho_{22}[((5-3\cos2\Delta\phi)\cos\theta^2+2\sin\theta^2] \nonumber \\
&&- \rho_{12}6\cos\theta\sin2\Delta\phi \nonumber \\
\rho_{12}' &=& 3\rho_{11}\sin2\Delta\phi\cos\theta' \nonumber \\
&&+ 6\rho_{12}(\cos\theta\cos\theta'\cos2\Delta\phi \nonumber \\
&&+ \sin\theta\sin\theta'\cos\Delta\phi) \\
&&+ 3\rho_{22}\cos\theta(-2\sin\theta\sin\theta'\sin\Delta\phi \nonumber \\
&&- \cos\theta\cos\theta'\sin2\Delta\phi) \nonumber \\
\rho_{22}' &=& \rho_{11}[(5-3\cos2\Delta\phi)\cos^2\theta'+2\sin^2\theta']
 \nonumber \\
&&+ \rho_{22}[(5+3\cos2\Delta\phi)\cos^2\theta\cos^2\theta' \nonumber \\
&&+ 2\cos^2\theta\sin^2\theta'
+12\cos\Delta\phi\cos\theta\cos\theta'\sin\theta\sin\theta' \nonumber\\
&&+ 2\cos^2\theta'\sin^2\theta+8\sin^2\theta\sin^2\theta')]  \nonumber \\
&&+ \rho_{12}(6\sin2\Delta\phi\cos^2\theta'\cos\theta \nonumber \\
&&+ 2\sin\Delta\phi\cos\theta'\sin\theta\sin\theta') \nonumber .
\end{eqnarray}

Finally, when the scattering is resonant with the transition 
${\rm S}_{1/2}\rightarrow {\rm P}_{1/2}$, 
then the density matrix for the scattered radiation is set to be
\begin{equation}
\rho_{11}'=\rho_{22}'={1\over2}, \quad \rho_{12}'=\rho_{21}'=0,
\end{equation}
which corresponds to the istropic and perfectly unpolarised radiation.

The angular distribution of the scattered radiation for an incident
photon with ${\bf{\hat k}}$ and $\rho$ is given by the
trace of $\rho'(\theta',\phi',\theta,\phi)$, as is shown by Lee et al. (1997).
Therefore, once the wavevector ${\bf{\hat k'}}$ of a scattered photon 
is chosen from $\rho'$ in accordance with the scattering type, the polarisation
is also determined at the same time. It is also noted that the degree of
polarisation is computed only after the new density matrix is normalised so
that it has a unit trace.

\begin{figure}
\centering
\includegraphics[width=9cm,angle=0]{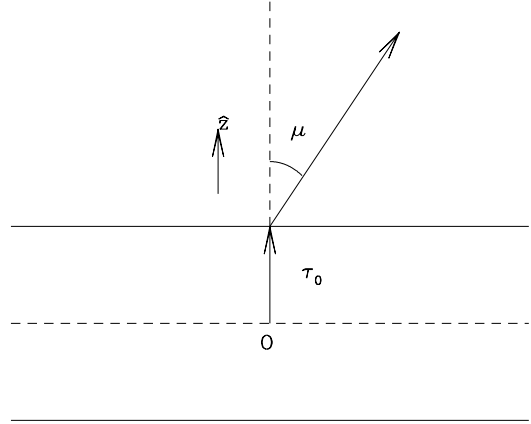}
\caption
{\small
Configuration of the model adopted in this work. 
We consider the static and uniform slab with the Ly$\alpha$ source 
at the origin denoted in the figure by $O$. The vertical thickness of
the slab is $2\tau_0$, and we define $\mu$ as the cosine
of the angle between the wave vector of the emergent photon
and the slab normal.}
\label{fig1}
\end{figure}

\section{Results}

\subsection{Degree of polarisation}

\begin{figure*}
\centering
\includegraphics[width=17cm,angle=0]{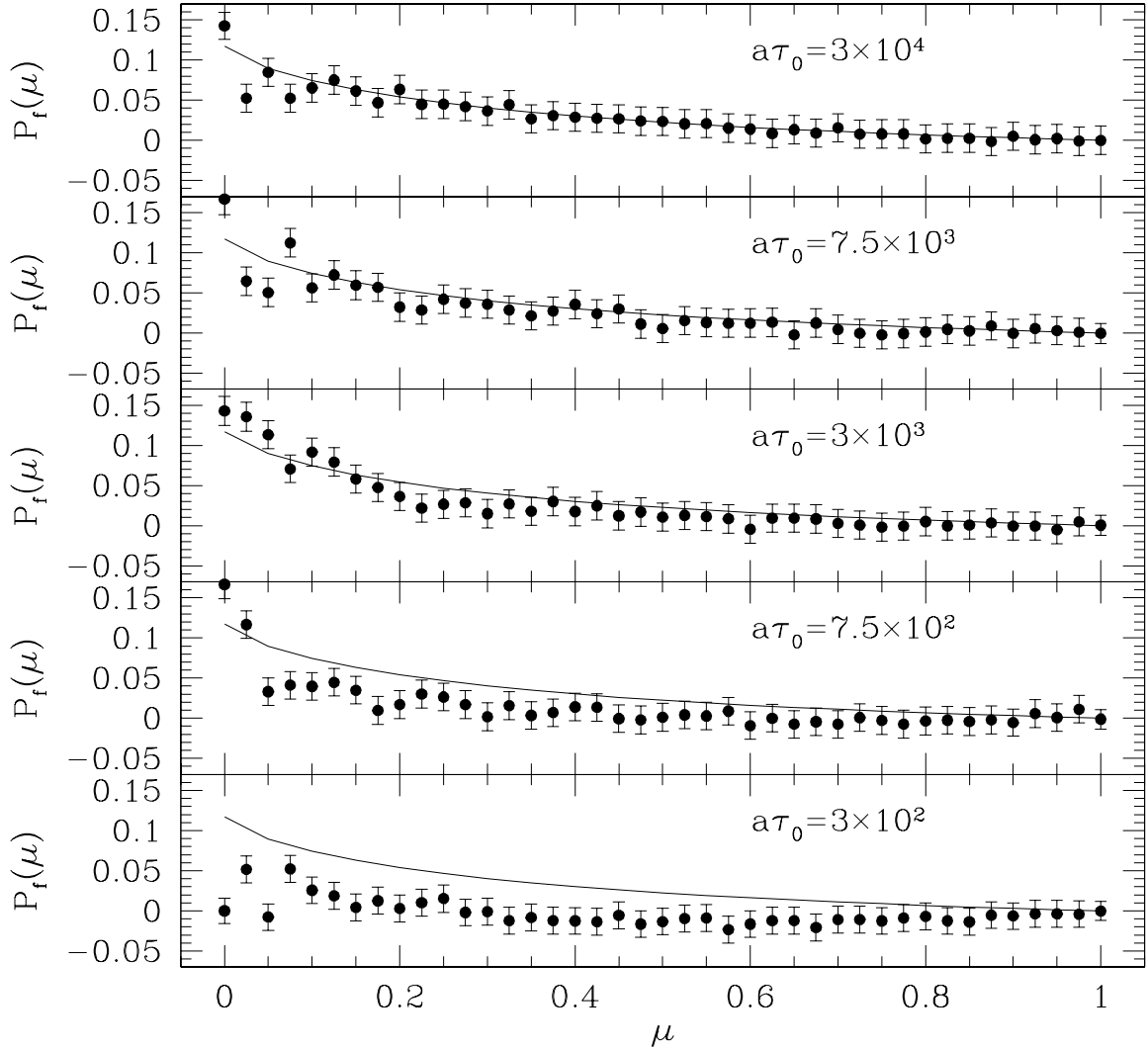}
\caption
{\small
Polarisation of the emergent Ly$\alpha$ photons originated from the 
unpolarised source located at the midplane of the slab illustrated in Fig.~1. 
We show five cases with different $a\tau_0$. The solid line in each box
represents the limiting behaviour for the Thomson scattering.}
\label{fig2}
\end{figure*}

We calculate the degree of linear polarisation of Ly$\alpha$ photons 
emerging from an optically thick slab with unpolarised sources 
distributed in its central plane. In Fig.~2, we show the 
linear degree of polarisation 
for various optical depths according to $\mu$ that represents the observer's
line of sight with respect to the slab normal. In the calculations, we consider
the cases where optical depths to the normal direction of the slab are
$\tau_0=2\times10^4, 5\times10^4,2\times10^5$, $5\times10^5$,
and $2\times10^6$. Since we fixed $a=1.49\times10^{-2}$,
$a\tau_0=3\times10^2, 8\times10^2, 3\times10^3$, and $8\times10^3$,
respectively. In the figure, the dots with $1\sigma$ error bars
represent our results and the solid lines represent the result for 
continuum photons transferred in a Thomson scattering semi-infinite
electron cloud (Chandrasekhar 1960). We can see in the figure that
the case with $a\tau_0 > 10^3$ shows good agreement
with the limiting case of the Thomson scattering.
We also see that the polarisation curves
gradually converge to that of the Thomson scattering 
with a very large optical thickness as $a\tau_0$ increases.

When $a\tau_0>10^3$, the Ly$\alpha$ transfer processes are dominated by
wing scatterings accompanied by the spatial transfer and Ly$\alpha$ photons 
experience a number of successive wing scatterings 
just before escape (Ahn et al. 2002).
The number of last successive wing scatterings is determined by the
scattering optical depth, $\tau_w$ at the characteristic frequency
of the emergent photons, given approximately by
\begin{equation}
\tau_w = { 1 \over \sqrt{\pi}} x_s,
\end{equation}
where $x_s=(a\tau_0)^{1/3}$. They also showed that these are
valid only for $a\tau_0\ge10^3$, when the diffusion
approximation can be safely applied.

For the cases considered in this section, the characteristic
wing optical depths are $\tau_w=3.8, 5.1, 8.1, 11.0, 17.5$, respectively.
According to Phillips \& M\'esz\'aros (1986), the limiting behaviour is almost
achieved when $\tau_e \ge 10$.  The polarisation of Ly$\alpha$ exhibits
the same limiting behaviour to that of the Thomson scattering
when $a\tau_0 \ge 10^3$ or $\tau_w \ge 10$.

\subsection{Spectropolarimetry}

\begin{figure}
\centering
\includegraphics[width=9.5cm,angle=0]{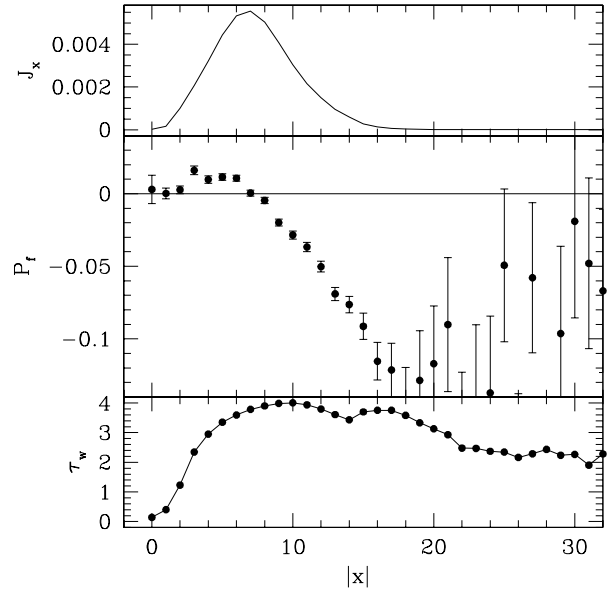}
\caption{\small
Simulated spectropolarimetry of Ly$\alpha$ emission for the unpolarised
source at the central plane of a slab with the vertical optical depth
from the centre of the plane $\tau_0=2\times 10^4$ and the Voigt parameter
$a=1.49\times 10^{-2}$.
The top panel shows the emergent flux, the middle panel shows
the degree of polarisation, and the bottom panel shows the effective wing 
optical depth $\tau^e_w=\sqrt{N_w}$ where $N_w$ is the number of successive 
wing scatterings just before escape.
All the quantities shown in the figure have been obtained after averaging 
over all emergent angles $\theta=\cos^{-1}\mu$.}
\label{fig3a}
\end{figure}

\begin{figure}
\centering
\includegraphics[width=9.5cm,angle=0]{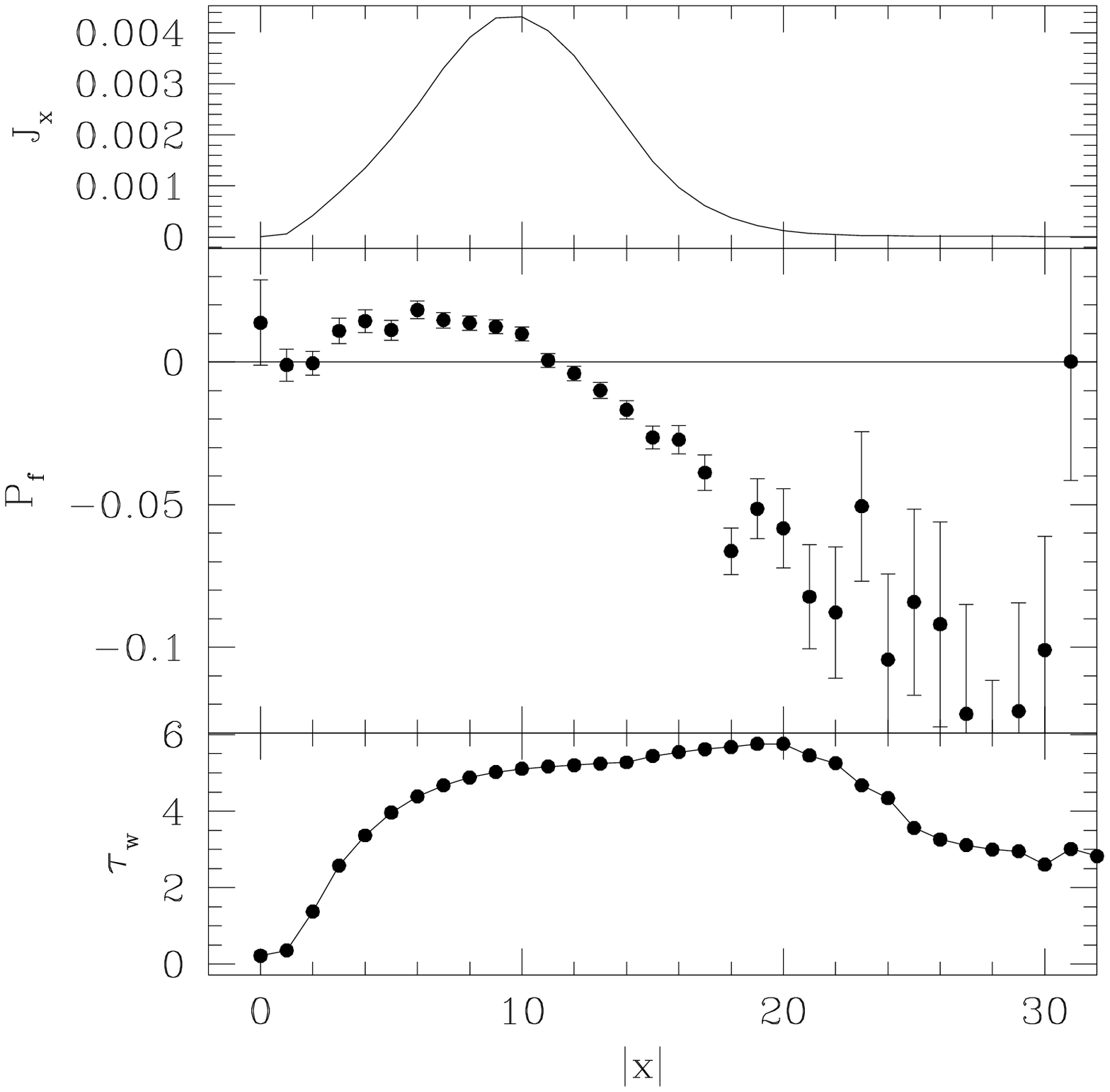}
\caption{\small
Simulated spectroplarimetry of Ly$\alpha$ from a slab with 
$\tau_0=5\times 10^4$. 
All other quantities represent the same as in Fig.~3.}
\label{fig3b}
\end{figure}

\begin{figure}
\centering
\includegraphics[width=9.5cm,angle=0]{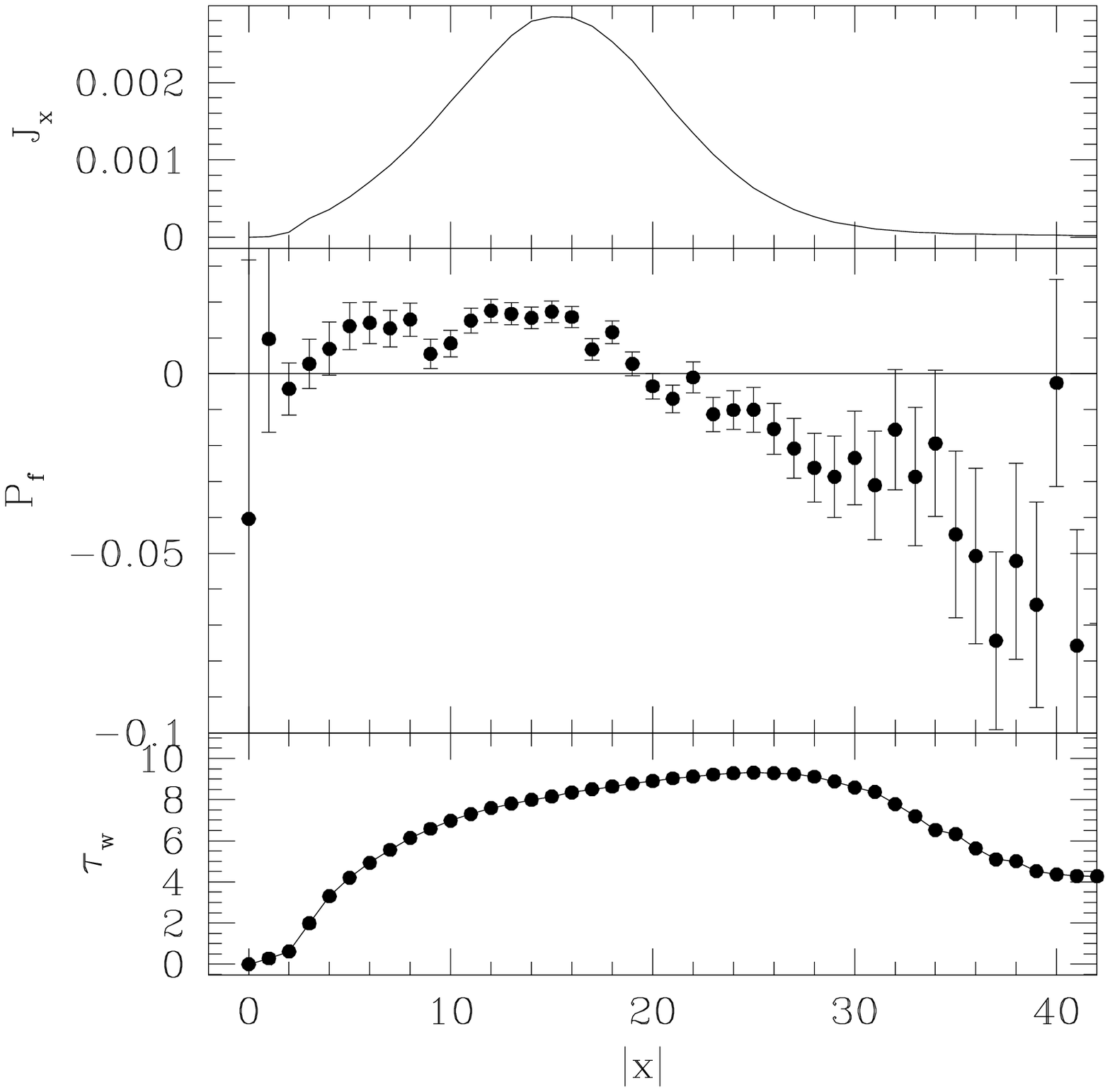}
\caption{\small 
Simulated spectroplarimetry of Ly$\alpha$ from a slab with 
$\tau_0=2\times 10^5$.
All other quantities represent the same as in Fig.~3.}
\label{fig3c}
\end{figure}

\begin{figure}
\centering
\includegraphics[width=9.5cm,angle=0]{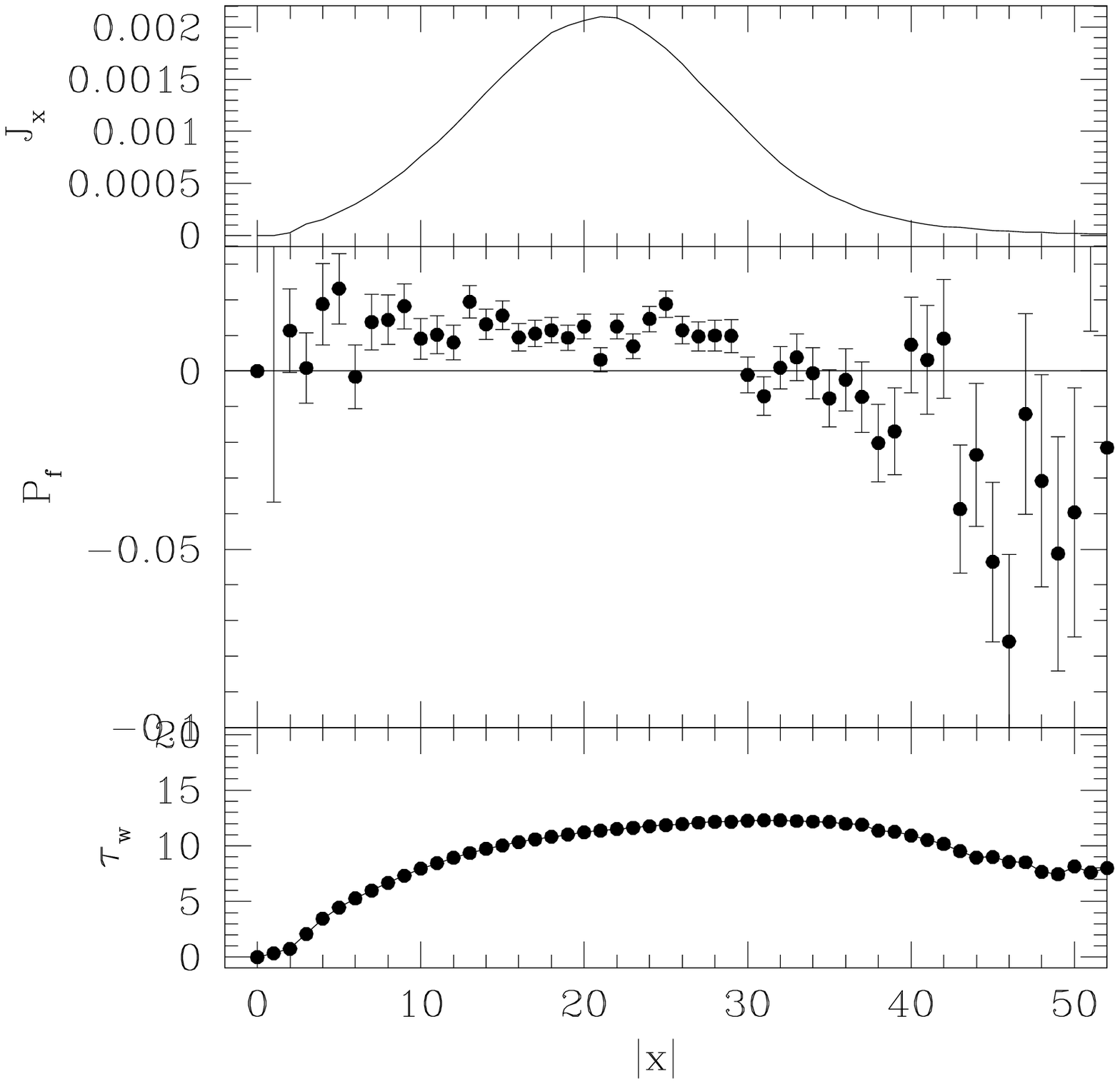}
\caption{\small
Simulated spectroplarimetry of Ly$\alpha$ from a slab with 
$\tau_0=5\times 10^5$.
All other quantities represent the same as in Fig.~3.}
\label{fig3d}
\end{figure}

\begin{figure}
\centering
\includegraphics[width=9.5cm,angle=0]{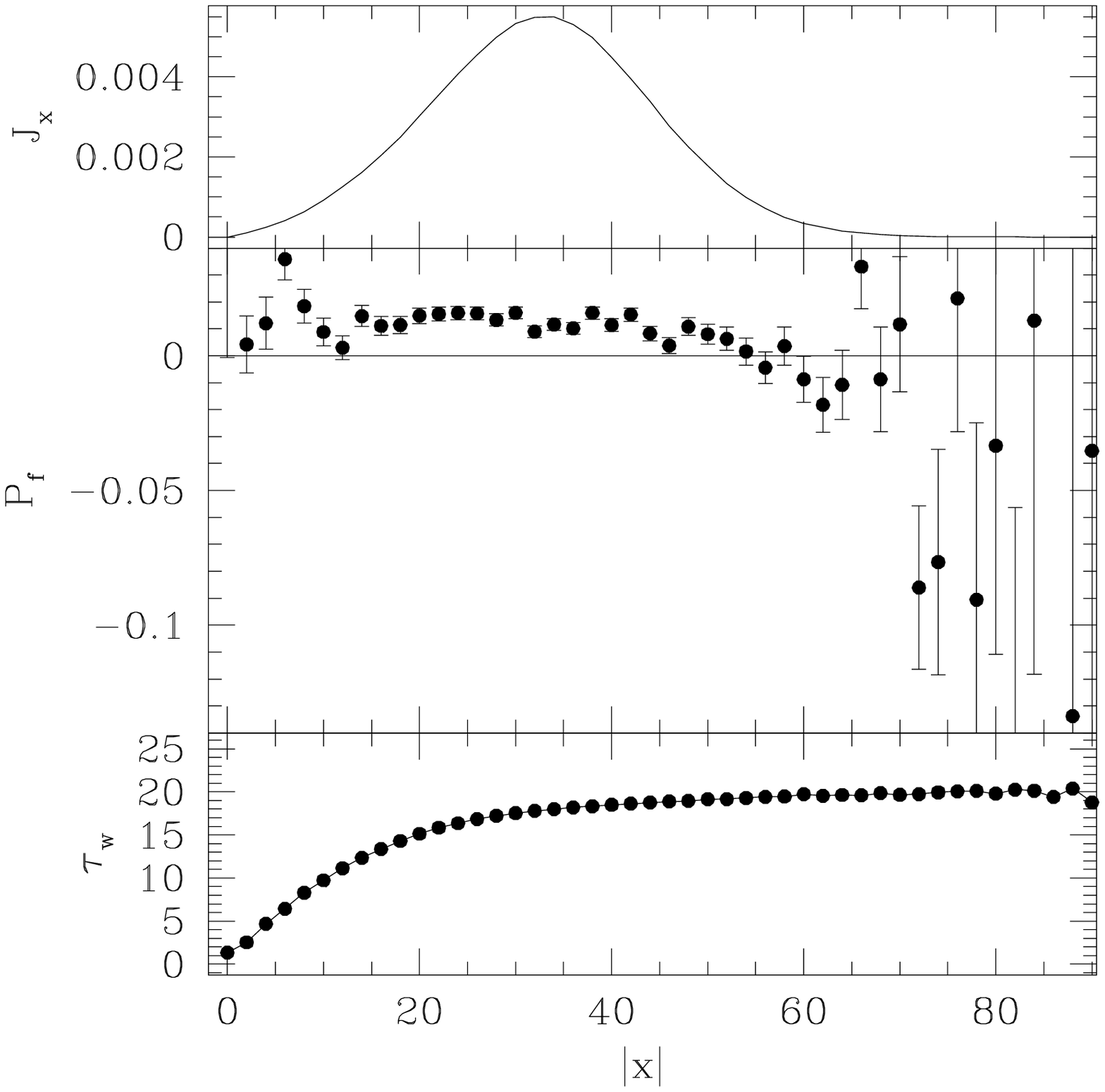}
\caption{\small
Simulated spectroplarimetry of Ly$\alpha$ from a slab with 
$\tau_0=2\times 10^6$.
All other quantities represent the same as in Fig.~3.}
\label{fig3e}
\end{figure}

We are interested in spectropolarimetric observations for many
astronomical objects. In Figs.~3-7 we show our results
of the accelerated Monte Carlo calculations
for the polarisation of Ly$\alpha$
emerging from an extremely thick hydrogen slab.
The total number of photons in each simulation is $1.6\times 10^6$,
and the frequency bin is set to be $\Delta x = 1$.
For $\tau_0 = 2\times 10^6$  we set $\Delta x = 2$  as an exception.
All the quantities shown in the figures have been obtained after averaging 
over all emergent angles $\theta=\cos^{-1}\mu$.

In the top panel of Figs.~3-7, we show the emergent fluxes,
in the middle panel the degree of polarisation $P_f$, and
in the bottom panel the characteristic optical depth, $\tau_w$.
Treating the radiative transfer in a Thomson scattering medium  as a 
random walk process, we convert $N_w$, the average number of successive 
wing scatterings just before escape, to the characteristic optical depth 
$\tau_w=\sqrt{N_w}$.

It is immediately seen that negligible polarisation develops near the 
line-centre.
This is because line photons with frequencies near the line-centre 
escape through a single longest flight after a large number of core
scatterings near the surface, where 
the local core scatterings isotropise the radiation field.
The core-wing boundary frequency is approximately given 
by Eq.(8) with $\tau_w=1$, or $x=\sqrt{\pi}$.
Hence our results in the figure show that $P_f=0$ at $0\le x \le \sqrt{\pi}$.

The second point is that the photons filling the near wing part 
of Ly$\alpha$ emission are positively polarised, which means the electric 
vectors of those escaping photons tend to lie in the direction perpendicular
to the slab normal. The degree of polarisation in this part is 
about $1\%$, which is
smaller than the maximum value $11.7\%$, because they are averaged over 
emergent angles. It is very notable that the degree of polarisation
becomes negative at far wing frequencies, which means
that the polarisation direction flips from perpendicular to parallel 
as the frequency shift $\Delta x$ from the line-centre increases.

In order for wing photons to acquire scattering numbers larger than 
$N_{w} =\tau_w^2$, they spatially diffuse much farther than $\tau_0$
in the stage of their final series of wing scatterings. 
According to Ahn et al. (2002), a line photon smears out towards 
the slab surface via wing scatterings, and therefore the photon diffuses 
into the grazing direction of the slab acquiring a sufficiently large 
frequency shift. This means that the scattering medium is 
effectively very thin in the normal direction 
from the view point of far wing scattering Ly$\alpha$ photons.
Therefore, the scattering plane associated with the escaping
photons nearly coincides with the slab plane, leading to the development
of polarisation in the direction parallel to the slab normal. 
This explains the negative polarisation in the far wing part illustrated 
in Figs.~3-7.

On the other hand, the near wing part of the emergent profiles is contributed
by photons with much larger wing scattering numbers. Therefore, a large
number of wing scatterings acompanied by spatial transfer preferentially
in the normal direction may cause
the electric vectors of the line photons to become perpendicular
to the slab normal. From this transfer process, we obtain positively 
polarised fluxes in accordance with our definition of polarisation direction.

Another point we can find in the figure is that the strength of the negative
polarisation in the far wing regime decreases as $a\tau_0$ increases.
This effect can be understood in terms of the beaming effect.
Ahn et al. (2002) investigated the beaming of Ly$\alpha$
photons for the cases of very large $a\tau_0$, which will also be shown
in the next subsection.
They showed that most photons escape from the extremely thick slab-like media 
favourably in the direction normal to the slab plane, as is the case 
with an electron scattering thick cloud (Phillips \& M\'esz\'aros 1986).
If this is the case, the fraction of Ly$\alpha$ photons
emergent in the direction parallel to the slab surface
decrease as $a\tau_0$ increases. 
As a result, their contribution to negative polarisation
diminishes, and eventually the degree of polarisation at the
far wings becomes larger and eventually positive.
Due to the beaming effect, the polarisation flip occurs
farther away from the line-centre as $a\tau_0$ increases.

The polarisation of Ly$\alpha$ in an anisotropically expanding slab
was investigated by Lee \& Ahn (1998).  They calculated the cases
with the line-centre optical depth $\tau_0 \le 10^5$
and the Voigt parameter $a=4.71\times 10^{-4}$.
Their results show that the polarisation flip
occurs near the line-centre, and as a result the emergent
peaks are negatively polarised on average over frequencies.
Because the bulk motion of the expanding slab enhances the photon escape, 
the polarization flip occurs at smaller optical depths than in the
static medium with the same neutral hydrogen column density. 

\subsection{Polarisation of the total flux}

\begin{figure}
\centering
\includegraphics[width=14cm,angle=0]{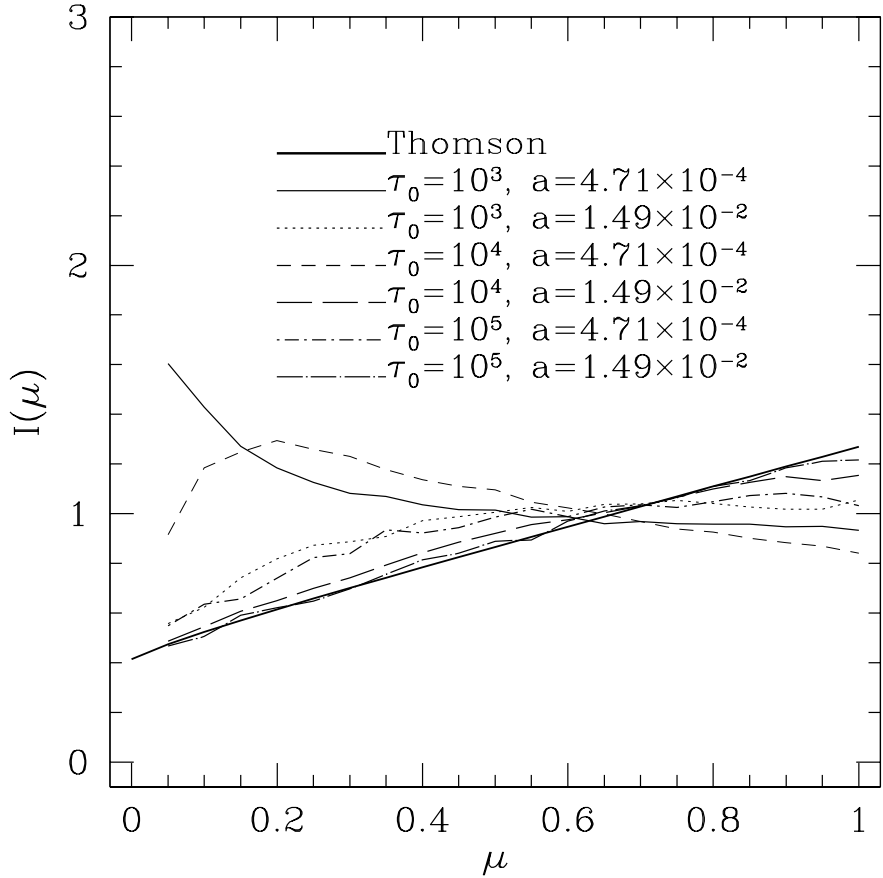}
\caption{\small
Directionality of emergent Ly$\alpha$ photons for various
optical depths $\tau_0$ and Voigt parameters $a$ of the scattering 
media. We define the directionality by the flux divided by $\mu$.
The solid thick line represents the limiting behaviour of directionality
for the Thomson scattering in an opaque electron cloud.
We note that the curves is convex upward when $a\tau_0\ll 10^3$
in the grazing direction.
As $a\tau_0$ is greater, the directionality gradually
converges to the curve for the Thomson scattering in a thick 
electron cloud.}
\label{fig4}
\end{figure}

Fig.~8 shows the directionality, $I(\mu)$, which is defined
by $F(\mu)/\mu$. Here $F(\mu)$ is the flux along $\mu$.
We note that this was also discussed in our previous paper
(Ahn et al. 2002).
Ahn et al. (2002) discovered that the Ly$\alpha$ limb brightening
occurs when $a\tau_0 \ll 10^3$, and that the Ly$\alpha$ limb
darkening appears when $a\tau_0>10^3$.

\begin{figure}
\centering
\includegraphics[width=14cm,angle=0]{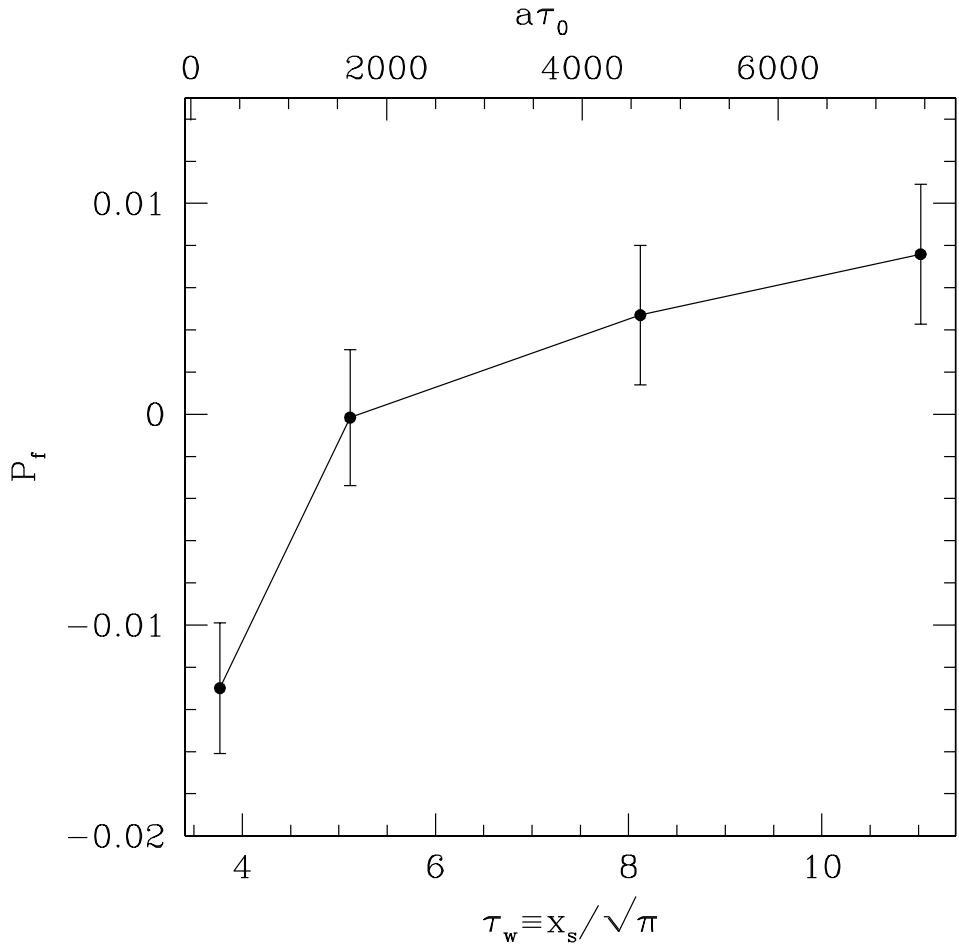}
\caption{\small
Degree of polarisation ($P_f$) when averaged over both
frequency and emergent angle.
We show the results for various optical depths
and Voigt parameters. $P_f$ has the negative value
for $a\tau_0 < 10^3$, while the positive values for $a\tau_0 > 10^3$.}
\label{fig5}
\end{figure}

In this subsection we show the polarisation ($P_t$)
of total emergent flux. This quantity is obtained from the definition
$P_t = \int P_f(\mu) I(\mu)\mu d\mu$, where $P_f(\mu)$ is shown
in Fig.~2 and $I(\mu)$ is shown in Fig.~8.

Fig.~9 shows the degree of polarisation as a function
of $a\tau_0$. Here $a\tau_0$ can be thought to be
a measure of relative importance between wing scatterings
and core scatterings in the line transfer.
The error in $P_t$ is simply propagated from $P_f(\mu)$.
The most important thing to note in Fig.~9 is that the polarisation
sign changes as $a\tau_0$.
For $a\tau_0 < 750$, the polarisation direction
is parallel to the slab normal, while
for $a\tau_0 > 750$ it is perpendiculat to it.

As we saw in the previous subsection 3.1, the case with
$\tau_0 = 2\times 10^4$ and $a=1.49\times10^{-2}$ has
the characteristic optical depth $\tau_w=3.8$.
We can understand the properties of Ly$\alpha$ wing
scatterings by comparing with the Thomson scattering
in an optically thick electron cloud with $\tau_e \approx \tau_w$,
because the scattering is described by the same Rayleigh phase function 
and spatial diffusion dominates the transfer process.

A close examination of the curves in Fig.~2 and Fig.~5 in 
Phillips \& M\'esz\'aros (1986)
and those in Fig.~2 and Fig.~8 in this work shows that
when $a\tau_0 <10^3$ the photon flux emergent in the grazing 
direction of the slab surface increases as $a\tau_0$ decreases.
On the other hand, the beaming of emergent Ly$\alpha$
is enhanced as $a\tau_0$ increases. 
As a result the critical frequency 
of the polarisation flip moves toward the larger value,
which we can see in Fig.~3-7.

\section{Summary }
We have investigated the linear polarisation of Ly$\alpha$
transferred in a static, dustless, and thick slab that is
uniformly filled with neutral hydrogen. The polarisation behavious
of the emergent Ly$\alpha$ emission can be understood qualitatively by
comparing with the polarisation behaviour of the Thomson scattered 
radiation. When the scattering medium is
extremely thick or $a\tau_0 > 10^3$, photons emerging
in the grazing direction are polarised perpendicular to the slab normal
with the degree up to $12\%$, which is the same value as that of 
the Thomson scattered radiation in a semi-infinite thick
medium (Chandrasekhar 1960).

The variation of linear polarisation with $a\tau_0$ 
also shows similar behaviour with that of 
the Thomson scattering (Phillips \& M\'esz\'aros 1986).
The linear polarisation of Ly$\alpha$ develops in the direction parallel
to the slab normal when $a\tau_0<10^3$, while it becomes perpendicular 
when $a\tau_0>10^3$.
Our simulated spectropolarimetry of Ly$\alpha$ shows the polarisation 
flip in the spectra. The linear polarisation
near the line centre is almost zero. The wing parts near the 
line centre are polarised in the direction perpendicular to the slab normal,
while the far wing parts are polarised in the direction parallel
to the slab normal. The zero polarisation around the resonance frequency is
caused by resonance scatterings just before escape which isotropises
the electric vector of the Ly$\alpha$ photons. The perpendicular polarisation
at the far wing parts is caused by Ly$\alpha$ photons preferentially
emerging in the grazing direction.

The Ly$\alpha$ photons constituting the near wing part of
the emergent peaks have large wing optical depths, and therefore 
they are beamed in the slab normal direction, which gives rise to 
positive polarisation. As $a\tau_0$ increases, the beaming effect 
prevents Ly$\alpha$ photons
with far wing frequencies from emerging into the grazing direction.
Therefore, the location of the polarisation flip in the Ly$\alpha$ line 
can be an important indicator of the value of $a\tau_0$.

\section*{Acknowledgments}

The authors thank Professor H. -M. Lee for helpful discussions
and a critical reading of the original version of the paper.
H.-W.L. gratefully acknowledges support from the Korea Research
Foundation Grant (KRF-2001-003-D00105).

\end{document}